\documentclass[12pt]{article}
\usepackage[authoryear]{natbib}
\usepackage[utf8]{inputenc}
\setcitestyle{round}
\usepackage{amsfonts, animate,subfigure, comment}
\usepackage{amssymb,amsthm,dsfont}
\usepackage[margin=1in]{geometry}
\usepackage{blkarray}
\usepackage{setspace}
\usepackage{amsmath}
\usepackage{url}
\usepackage{graphics}
\usepackage{graphicx,color}
\usepackage{multirow}

\usepackage[colorlinks, urlcolor = blue, citecolor = blue]{hyperref}
\usepackage[ruled]{algorithm2e}
\usepackage{bm}
\usepackage[format=hang,labelfont=bf,font={small,it}]{caption}
\usepackage[capposition=below]{floatrow}
\usepackage{paralist}
\usepackage{soul}
\usepackage{tikz}

\usepackage{color}

\usepackage[rightbars, color]{changebar}
\cbcolor{blue}
\newcommand{\beq}{\begin{equation}}
\newcommand{\eeq}{\end{equation}}

\newtheoremstyle{case}{}{}{}{}{}{:}{ }{}
\theoremstyle{case}

\begin{document}
\renewcommand{\thefootnote}{\arabic{footnote}}
\vspace{-3cm}
\title{
    \fontencoding{T1}
  \fontfamily{timesnewroman}
  \fontseries{m}
  \fontshape{it}
  \fontsize{22}{22}
 \selectfont Anomaly Detection in Large Scale Networks with Latent Space Models\\ 
 }
\normalfont 
\vspace{1cm}
\author{Wesley Lee \\
\vspace{.1cm}
University of Washington \and
Tyler H. McCormick\thanks{Contact email: tylermc@uw.edu} \\
\vspace{.1cm}
University of Washington
\and
\and
Joshua Neil \\
\vspace{.1cm}
Microsoft
\and
Cole Sodja \\
\vspace{.1cm}
Microsoft
\and
Yanran Cui \\
\vspace{.1cm}
University of Washington
}
\date{}
\maketitle
\vspace{-.4cm}

\begin{abstract}

We develop a real-time anomaly detection method for directed activity on large, sparse networks. We model the propensity for future activity using a dynamic logistic model with interaction terms for sender- and receiver-specific latent factors in addition to sender- and receiver-specific popularity scores; deviations from this underlying model constitute potential anomalies. Latent nodal attributes are estimated via a variational Bayesian approach and may change over time, representing natural shifts in network activity. Estimation is augmented with a case-control approximation to take advantage of the sparsity of the network and reduces computational complexity from $O(N^2)$ to $O(E)$, where $N$ is the number of nodes and $E$ is the number of observed edges. We run our algorithm on network event records collected from an enterprise network of over 25,000 computers and are able to identify a red team attack with half the detection rate required of the model without latent interaction terms.  Code to replicate the results in this paper are available at \url{https://github.com/thmccormick/replicate-latent-space-networks-anomaly-det.git}. 
\end{abstract}

\doublespacing

\section{Introduction} \label{section1:Introduction}

The near ubiquity of reliable and cost-effective telecommunications technology means that even small or medium size organizations maintain enterprise networks with thousands of interconnected devices.  The ``Internet of Things (IoT),'' bringing with it smart homes, streets, cars, and offices, has further increased the number and type of networked devices.  Modern networks now consist of a mix between devices where electronic communication is a primary function (e.g. computers, servers, or smart speakers) and devices that are intended for another purpose but maintain network connectivity for convenience or remote access (e.g. home appliances or cars).

These expanding enterprise networks provide unparalleled access and convenience for device users, but also increase potential points of vulnerability for cyberattacks.  {Various statistical methods have been adapted for network surveillance, including control chart and hypothesis testing methods, Bayesian methods, scan methods, and time series models (see \citet{woodall2017overview} or \citet{jeske2018survei} as examples).
\citet{yu2018node} proposed the adoption of the compositional $T^2$ control chart to the social network setting and used it for monitoring each network community separately. \citet{priebe2005enron} proposed a general scan statistics method for anomaly detection and \citet{zhao2018temporal} applied it on the degree corrected stochastic block (DCSBM)
model. Some recent advances also combined these methods to produce improved anomaly detection results. \citet{leto2015detect} utilized a hierachical network model and proposed a prior with Beta distribution. They constructed the test statistic based on the posterior for detection change point and found their resulting method is more accurate than the methods based on converting the network sequence into a time series of scalar values
(mean degree, mean geodesic distance, mean local
clustering coefficient). \citet{wang2017hierarchical} further extended this method via graph contraction in order to distinguish between intra-community and inter-community evolution. However, as the scale of networks increase tremendously, more flexible methods that can perform online anomaly detection are in greater demand. Considering this recent work, a salient feature of the enterprise scale security context is that, in most cases, the number of interactions is large so only methods with a very low false positive rate can be implemented realistically in practice.  Otherwise, even a method that performs will overall will overwhelm a team tasked with following up on attack behavior.}

In this paper, we present a scaleable statistical approach for identifying anomalous behavior in enterprise networks.  We use latent space models \citep{hoff2002latent} to parsimoniously represent the structure of the connectivity graph between devices, capturing the likely complex dependence structure.  This approach provides a richer baseline description of network behavior than current approaches that, for computational reasons, focus on either single device behavior or small clusters of devices.  We use a combination of variational and case-control approximations to ensure that our approach can both represent rich network structure and be implemented on the large-scale networks. {Our approach provides scores representing deviations from this baseline behavior.  These scores can then either be used to develop a threshold to produce an alarm or, as we focus on primarily in this paper, be used to rank potentially anomalous activities for follow-up investigation.} 

Our approach simultaneously addresses three key challenges which are critical for anomaly detection in cybersecurity settings.  Our approach (i) uses a baseline model for behavior on the network that incorporates higher order dependence between networked devices, (ii) includes a scaleable computational algorithm that leverages the sparsity in enterprise networks and (iii) captures dynamics through an efficient online updating scheme.

First, we use an anomaly detection framework where we build a model that represents normal behavior and use deviations from this model to flag potential anomalies (see e.g. \citep{neil2013detection} for other examples that use this setup). Each pair of devices, or dyad, has a propensity to interact in a given time period.  If observed device interactions are very unlikely under the baseline model, then an anomaly is registered for further investigation.   
The baseline we propose incorporates higher order network properties using a latent position model \citep{hoff2002latent}.  These models represent the likelihood of two devices interacting based on distance in an unobserved geometric space.  The closer the estimated positions of the two devices, the more likely they are to interact.  The latent distance model captures higher order features of the graph, such as the propensity to form triangles, through the geometry of the latent space.  Our approach is in contrast to existing baseline models which, for computational reasons, avoid complex interaction terms, opting for simple sender- and receiver-specific popularity terms \citep{neil2013towards} or clustering nodes and modeling interactions at a cluster level \citep{metelli2016model}.  While an approach that relies on individual device effects will be effective for capturing some types of attacks (e.g. if an attacker uses an infected device in a way that dramatically increases the activity of that device), it will fail to detect more sophisticated attacks that change the pattern of device behavior rather than simply the volume of interaction. 

Second, we develop a strategy for online computation that leverages the sparsity in enterprise networks.  As alluded to above, computation in network models depends not on the scale of devices, but on the scale of device pairs, or dyads. If $N$ is the number of devices, then there are {approximately} $N^2$ possible dyads.  A system with 25,000 devices, for example, will have over 600 million dyads, making any algorithm that requires evaluating the full likelihood via summation of over device pairs completely infeasible.  This scale also precludes Bayesian computation using Markov chain Monte Carlo (MCMC), as is typically done for network models{ \citep{snijders2002mcmc, golightly2011bayes, krause2009mcmc}}.  We leverage the variational message passing algorithm described in \citet{minka2005divergence} with a case control approximation inspired by \citet{raftery2012fast} in order to reduce the computation cost from $O(N^2)$ to $O(E)$, where $E \ll N^2$ is the number of observed edges in the network.  Since the computational complexity scales with the number of realized connections, rather than the number of possible ones, the algorithm will be particularly efficient for sparse graphs.  Sparsity is a common feature in enterprise networks{ \citep{jia2019sparsity}}. For example, most end-user machines do not talk to each other, rather they communicate with servers, vastly reducing the number of dyads. In the data we use in our empirical evaluation, for example, an average of 0.02\% possible dyads communicate in any four hour period.

Our method has similarities to existing dynamic network models using latent projections and to previous work on variational inference for network models. \citet{sewell2015latent} and  \citet{durante2016locally} both propose models with temporal dynamics, but are difficult to scale to graphs of the size we encounter in enterprise networks. \citet{variational2013salter} also propose a variational algorithm in the static latent position model \citep{hoff2002latent}.  Their approach is based on minimizing KL divergence, finding substantial computational gains over a comparable MCMC even before using the case-control approximation from \citet{raftery2012fast}. The primary expectation required in the algorithm is inherently intractable, and they proceed via a series of Taylor series expansions in order to reach a tractable expression. In contrast, we choose to adopt a variational approach minimizing a different divergence metric but resulting in a tractable, analytic set of updating equations. ~\citet{sewell2017latent} also propose a scaleable computational algorithm for dynamic networks, but focus on detecting community structure rather than identifying anomalous behavior.

Third, we propose an efficient dynamic updating procedure to capture variations in behavior over time. Taking advantage of the parametric form of the variational approximation to the posterior, we allow our parameters to update with discrete time dynamics via Gaussian random walks, adopting the autotuning procedure from \citet{mccormick2012dynamic} in order to flexibly adjust the amount of additional variation introduced at each time step.
In addition to their autotuning procedure, \citet{mccormick2012dynamic} propose a general purpose algorithm for estimating dynamic logistic regression models. However, their approach jointly updates the logistic parameters via Newton's method and requires inversion of the corresponding Hessian. When the number of parameters in the model is large, such as when each node has a specific popularity term, this process becomes infeasible.

We evaluate our method using the Netflow activity data collected by the Los Alamos National Laboratory (LANL) on their enterprise network for a period of 89 days \citep{turcotte2018unified}. Each record consists of directed communication between network devices, and network activity is logged between over 25,000 devices over these 89 days. In cybersecurity, a major objective is to flag potential intrusions on the network. The detection of these invaders is time sensitive, reflecting a desire to prevent further intrusion when possible.  The LANL data also contain a so-called ``red team'' attack, which is a simulated cyberattack that mimics tactics used by actual attackers.  The ``red team'' attack provides a ground truth event to use for benchmarking discovery.  The data are available at \url{https://csr.lanl.gov/data/2017.html}.

The remainder of the paper is organized as follows. In Section \ref{section1: latent space model} we describe the static bilinear effects model. In Section \ref{section1:vi} we present the variational message passing algorithm for the static model as well as the case-control modification. Section \ref{section1:dynamics model} adapts the model and algorithm to a dynamic setting, and Section \ref{section1:anomaly detection} describes anomaly detection after estimation is complete. In Section \ref{section1:simulation} we demonstrate the performance of our algorithm in a simulation study, in Section \ref{section1:LANL data} we apply our algorithm to the LANL computer network data, and in Section \ref{section1:Discussion} we conclude.

\section{Dynamic Latent Space Models} \label{section1: Model Section}

In this section we present our model and computation strategy.  We first present a network model and computational approach for static graphs and then describe how we incorporate temporal dynamics.

\subsection{Bilinear Mixed-Effects Model} \label{section1: latent space model}

To model baseline behavior at a given time point we use the logistic specification of the bilinear mixed-effects model \citep{hoff2003bilinear}. Letting $y_{i,j}$ indicate the presence of directed activity from sender $i$ to receiver $j$ (e.g. a message passed from a computer to a networked printer), under this model
\begin{equation}
    y_{i,j} = \text{Bernoulli}(p_{i,j}),
\label{eq1:logit_eq}
\end{equation}
where
\begin{equation}
    \text{logit}(p_{i,j}) = \mu + \alpha_i + \beta_j + u_i^Tv_j.
\label{eq1:logit_bmem}
\end{equation}
$\mu$ controls the overall sparsity of the network, while $\alpha_i$ and $\beta_j$ represent sender- and receiver-specific popularity terms and $u_i$ and $v_j$ are $d$-dimensional sender and receiver-specific latent factors.  The interaction term $u_i^Tv_j$ captures the affinity between $i$ and $j$, and can be interpreted as the additional propensity for senders with certain latent characteristics to interact with receivers with other certain latent characteristics over the baseline propensity implied by their respective popularities. 

We complete our Bayesian model by introducing independent Gaussian priors for each of the parameters:\\
\begin{center}
    \begin{tabular}{c c}
         $\alpha_i  \sim N(0,\sigma_{\alpha})$\hspace{20pt} &\hspace{20pt} $\beta_j  \sim N(0,\sigma_{\beta})$  \\
       $ u_i  \sim N(0,\Sigma_u)$\hspace{20pt} & \hspace{20pt}
        $v_j  \sim N(0,\Sigma_v)$\\
        $\mu  \sim N(0,\sigma_\mu)$.&
    \end{tabular}
\end{center}
Lastly, we use $N$ to denote the number of nodes in the network and the $N \times N$ matrix $\textbf{Y}$ to denote all directed activity in the network. Without loss of generality, we assume the number of senders and the number of receivers in the network are equal.

\subsection{Variational Inference} \label{section1:vi}

Let $\theta\equiv\{\mu,\alpha_i,...\beta_j,...u_i,...v_j...\}$ denote the set of latent variables in the bilinear mixed-effects model. Given the large number of parameters, we wish to construct a parsimonious representation of the posterior $p(\theta|\textbf{Y})$. For example, the posterior covariance between the sender- and receiver-specific popularity terms would require the storage of a $N\times N$ matrix. To this end, we focus on learning a fully-factorized approximation $q(\theta)$ to the posterior, with independent terms for each latent variable. Specifically,
\begin{equation}
q(\theta) = q(\mu) \prod_i q(\alpha_i) q(u_i) \prod_j q(\beta_j) q(v_j),
\label{eq1:post approximation}
\end{equation}
where each marginal term is modeled with a Gaussian (with $d\times d$ covariance matrices for each of the latent factor terms). Representing the posterior of each latent variable with an independent Gaussian leads to a storage complexity of $O(N)$ and will also help facilitate the introduction of temporal dynamics in the following section.

Inference proceeds as a direct application of power expectation propagation (Power EP) \citep{minka2005divergence} and takes the form of a message passing algorithm. We describe the algorithm in detail below but omit some of the theoretical basis provided in \citep{minka2005divergence}. First, we recast the posterior as a product of factors, where each factor is either a dyadic observation or a prior over a latent variable.
\begin{align}
p\left(\theta|\textbf{Y}\right) & \propto p\left(\textbf{Y}|\theta\right)p\left(\theta\right)\\
 & \propto\prod_{\left(i,j\right)}f_{\left(i,j\right)}\left(\theta\right)
\label{eq1:factorized posterior}
\end{align}
We use $\left(i,j\right)$ to index the factor denoting the directed dyad $i{\rightarrow}j$, with $\left(0,k\right)$ for the factor denoting the prior over the $k$th latent variable $\theta_k$. We arbitrarily index the set of latent variables in our model with $k$ for notational simplicity for contexts in which the differences between the various latent variables are unimportant. {We cast our fully-factorized approximation, $q$, in light of the same factors (see equation \ref{eq1:post approximation}) such that the approximation to the posterior for $\theta_k$ is the product of messages from each of the factors $(i,j)$ to $\theta_k$:

\begin{equation*}
q\left(\theta\right)=\prod_{\left(i,j\right)}\tilde{f}_{\left(i,j\right)}\left(\theta\right).
\label{eq1:factorized approximation q}
\end{equation*} 
Each factor will be approximated by a member of exponential family and, since the exponential family is closed under multiplication, the product of the approximations of the factors will provide an approximation of $p$. 

The difficulty here, however, is that the best approximation of each factor depends on the rest of the network. We address this using an iterative message passing procedure. Under power expectation propagation, inference proceeds by iteratively selecting a single factor $(i,j)$ and updating the approximation by minimizing the $\boldsymbol{\alpha}$-divergence between $f_{(i,j)}\prod_{(i^\prime,j^\prime)\neq(i,j)} \tilde{f}_{\left(i^\prime,j^\prime\right)}$ and $\tilde{f}_{(i,j)}\prod_{(i^\prime,j^\prime)\neq(i,j)} \tilde{f}_{\left(i^\prime,j^\prime\right)}$.This local $\boldsymbol{\alpha}$-divergence approximates the minimization of the global $\boldsymbol{\alpha}$-divergence between the posterior $p$ and approximation $q$ under the assumption that the other factors $f_{(i^\prime,j^\prime)}$ are well-approximated by $\tilde{f}_{(i^\prime,j^\prime)}$. Here the $\boldsymbol{\alpha}$-divergence is defined by \citep{minka2005divergence} as 
\begin{equation*}
D_{\alpha}\left(p\parallel q\right)=\frac{\int\alpha p(\theta)+(1-\alpha)q(\theta)-p(\theta)^{\alpha}q(\theta)^{1-\alpha}d\theta}{\alpha(1-\alpha)}.
\label{eq1:alpha divergence}
\end{equation*}
As $\alpha \rightarrow 0$ or $\alpha \rightarrow 1$, the $\boldsymbol{\alpha}$-divergence converges to the traditional KL divergence.}
{Based on the updating procedure, the algorithm can be interpreted as message passing between the factors. The approximation of $f_{(i,j)}$ can be treated as the message $f_{(i,j)}$ sends to the rest of the network. Correspondingly, $\prod_{(i^\prime,j^\prime)\neq(i,j)} \tilde{f}_{\left(i^\prime,j^\prime\right)}$ can be treated as the collection of messages that factor $f_{(i,j)}$ receives which represents the behavior of the rest of the network. Considering the interpretation of $\tilde{f}_{(i,j)}$, the factor approximations are fully factorized into messages as
\begin{equation*}
f_{\left(i,j\right)}\left(\theta\right) \approx \tilde{f}_{\left(i,j\right)}\left(\theta\right) \equiv \prod_k m_{\left(i,j\right)\rightarrow\theta_k}\left(\theta_k\right).
\label{eq1:approximated factor}
\end{equation*}
which yields the full approximation 
\begin{equation*}
q\left(\theta_{k}\right)=\prod_{\left(i,j\right)}m_{\left(i,j\right)\rightarrow\theta_{k}}\left(\theta_{k}\right).
\label{eq1:factorized approximation}
\end{equation*}}

Each message can be conceptualized as the contribution of a single factor to the posterior of a single variable. If a variable is not involved with a given factor (e.g. the sender popularity $\alpha_j$ when considering the factor $i{\rightarrow}j$ with sender $i$ and receiver $j$), the message is uniform and provides no contribution to the posterior. 
As our approximation $q$ is a product of Gaussian densities, we take the messages to be unnormalized Gaussian densities, noting that these densities are closed under multiplication and we can implicitly rescale $q(\theta_k)$ to be a (normalized) Gaussian density after every iteration.

{In order to minimize local $\boldsymbol{\alpha}$-divergence, based on \citep{minka2005divergence} Theorem 3, this problem can be converted to finding the KL projection of $p(\theta)^{\alpha}q(\theta)^{1-\alpha}$ to the family of Gaussian densities. With the full factorization of $q$ into product of messages and the discussion of section 4.1 in \citep{minka2005divergence}, the update step for variable $\theta_k$ from factor $(i,j)$ is given by}

\begin{gather}
q^{\prime}\left(\theta_{k}\right) =\text{proj}\left[q\left(\theta_{k}\right)m_{\left(i,j\right)\rightarrow\theta_{k}}^{-\alpha}\left(\theta_{k}\right)\int_{\theta\backslash\theta_{k}}f_{\left(i,j\right)}^{\alpha}\left(\theta\right)\prod_{\theta\backslash\theta_{k}}q\left(\theta\right)m_{\left(i,j\right)\rightarrow\theta}^{-\alpha}\left(\theta\right)d\theta\right], \label{eq1:update step}\\
q\left(\theta_{k}\right)^{\text{new}} =q\left(\theta_{k}\right)^{\epsilon}q^{\prime}\left(\theta_{k}\right)^{1-\epsilon},\label{eq1:update step 2}\\
m_{(i,j)\rightarrow\theta_{k}}\left(\theta_{k}\right)^{\text{new}} = \frac{q\left(\theta_{k}\right)^{\text{new}} m_{(i,j)\rightarrow\theta_{k}}\left(\theta_{k}\right)}{q\left(\theta_{k}\right)}.
\label{eq1:message update step}
\end{gather}
where {$\int_{\theta\backslash\theta_k}$ means the integral over all parameters except for $\theta_k$} and $\text{proj}\left[p\right]=\text{argmin}_{q}\text{KL}\left(p||q\right)$ denotes KL projection to the family of Gaussian densities (matching the mean and variance of $p$) and $\epsilon$ is a damping factor to aid with the convergence of the algorithm. Define $g(\theta_l) \equiv q\left(\theta_l\right)m_{\left(i,j\right)\rightarrow\theta_l}^{-\alpha}\left(\theta_l\right)$ and note $g(\theta_l)$ has the form of a Gaussian density, which we can take to be normalized. Then equation (\ref{eq1:update step}) can be written as
\begin{equation}
q^{\prime}\left(\theta_{k}\right) =\text{proj}\left[g\left(\theta_{k}\right)E_{g,\theta\backslash\theta_k}\left[f_{\left(i,j\right)}^{\alpha}\left(\theta\right)\right]\right].
\label{eq1:simple update step}
\end{equation}
One particular strength of the Power EP approach is the ability to choose $\boldsymbol{\alpha}$ such that evaluating the above expectations is tractable. The choice of $\boldsymbol{\alpha}$ also affects the shape of the approximation $q$ relative to $p$. Minka \citep{minka2005divergence} notes the choice of $\boldsymbol{\alpha}=-1$ puts greater emphasis on concentrating the mass of $q$ inside higher density areas of the $p$ (as opposed to ``covering" the posterior) and can lead $q$ to understate the variability in the posterior. For the logistic likelihood, the choice $\alpha=-1$, is particularly compelling:
\begin{align}
E_{g,\theta\backslash\theta_k}\left[f_{\left(i,j\right)}^{-1}\left(\theta\right)\right] & =E_{g,\theta\backslash\theta_k}\left[1+\text{exp}\left(-y_{ij}\left(\mu+\alpha_{i}+\beta_{j}+u_{i}^{T}v_{j}\right)\right)\right]\\
 & =1+E_{g,\theta\backslash\theta_k}\left[\text{exp}\left(-y_{ij}\mu\right)\text{exp}\left(-y_{ij}\alpha_{i}\right)\text{exp}\left(-y_{ij}\beta_{j}\right)\text{exp}\left(-y_{ij}u_{i}^{T}v_{j}\right)\right]
\label{eq1:alpha choice}
\end{align}
where the final expectation factors over each term. There are three sets of expectations to evaluate: $E_{g,\mu}\left[\exp\left(-y_{ij}\mu\right)\right]$
(and equivalent expressions for the other univariate parameters), $E_{g,v_{j}}\left[\exp\left(-y_{ij}u_{i}^{T}v_{j}\right)\right]$,
and $E_{g,u_{i},v_{j}}\left[\exp\left(-y_{ij}u_{i}^{T}v_{j}\right)\right]$.
The first two can be evaluated directly from the moment generating
functions for the univariate and multivariate Gaussian distribution,
and the last can be evaluated using the independence between the distributions
over $u_{i}$ and $v_{j}$ with complete the square techniques. Using $\mu_{\theta_k}$ and $\sigma_{\theta_k}$ or $\Sigma_{\theta_k}$ when appropriate to denote the mean and variance of $g(\theta_k)$:
\begin{align}
E_{g,\mu}\left[\exp\left(-y_{ij}\mu\right)\right] & =\text{exp}\left(-y_{ij}\mu_{\mu}+\frac{1}{2}\sigma_{\mu}^{2}\right),\\
E_{g,v_{j}}\left[\exp\left(-y_{ij}u_{i}^{T}v_{j}\right)\right] & =\text{exp}\left(-y_{ij}\mu_{v_{j}}^{T}u_{i}+\frac{1}{2}u_{i}^{T}\Sigma_{v_{j}}u_{i}\right),\\
E_{g,u_{i},v_{j}}\left[\exp\left(-y_{ij}u_{i}^{T}v_{j}\right)\right] & =E_{g,u_{i}}\text{\ensuremath{\left[\text{exp}\left(-y_{ij}\mu_{v_{j}}^{T}u_{i}+\frac{1}{2}u_{i}^{T}\Sigma_{v_{j}}u_{i}\right)\right]}}\\
 & =\text{det}\left(\Sigma_{v_{j}}^{-1}-\Sigma_{u_{i}}\right)^{-\frac{1}{2}}\text{det}\left(\Sigma_{v_{j}}\right)^{-\frac{1}{2}}\text{exp}\left(-\frac{1}{2}\mu_{v_{j}}^{T}\Sigma_{v_{j}}^{-1}\mu_{v_{j}}\right)\\
 & \,\,\times\text{exp}\left(\frac{1}{2}\left(\mu_{u_{i}}+\Sigma_{v_{j}}^{-1}\mu_{v_{j}}\right)^{T}\left(\Sigma_{v_{j}}^{-1}-\Sigma_{u_{i}}\right)^{-1}\left(\mu_{u_{i}}+\Sigma_{v_{j}}^{-1}\mu_{v_{j}}\right)\right).
 \label{eq1:expectations}
\end{align}
Completing the square, used in the last equation, relies on $\Sigma_{v_{j}}^{-1}-\Sigma_{u_{i}}$ being a positive definite matrix in order for the resulting density to be a multivariate normal distribution. 

We focus on the update steps for $\alpha_i$ and $u_i$, noting the symmetry in equation (\ref{eq1:alpha choice}) with respect to $\mu$, $\alpha_i$, and $\beta_j$, and similarly for $u_i$ and $v_j$, implies the corresponding update steps can be obtained by swapping the positions of the relevant variables. The updates take the form
\begin{align}
q^{\prime}\left(\alpha_{i}\right) & =\text{proj}\left[g\left(\alpha_{i}\right)\left(1+c_1\exp\left(-y_{ij}\alpha_{i}\right)\right)\right],\\
q^{\prime}\left(u_{i}\right) & =\text{proj}\left[g\left(\alpha_{i}\right)\left(1+c_2\text{exp}\left(-y_{ij}\mu_{v_{j}}^{T}u_{i}+\frac{1}{2}u_{i}^{T}\Sigma_{v_{j}}u_{i}\right)\right)\right],
\label{eq1:update with constants}
\end{align}
with
\begin{align*}
c_1 & = E_{\mu}\left[\exp\left(-y_{ij}\mu\right)\right]E_{\beta_{j}}\left[\exp\left(-y_{ij}\beta_{j}\right)\right]E_{u_{i},v_{j}}\left[\exp\left(-y_{ij}u_{i}^{T}v_{j}\right)\right],\\
c_2 & = E_{\mu}\left[\exp\left(-y_{ij}\mu\right)\right]E_{\alpha_{i}}\left[\exp\left(-y_{ij}\alpha_{i}\right)\right]E_{\beta_{j}}\left[\exp\left(-y_{ij}\beta_{j}\right)\right].
\end{align*}
These densities can be represented as a linear combination of two
Gaussian densities:
\begin{align}
q^{\prime}\left(\alpha_{i}\right) & =\text{proj}\left[N\left(\alpha_{i};\mu_{\alpha_{i}},\sigma_{\alpha_{i}}^{2}\right)+c N\left(\alpha_{i};\mu_{\alpha_{i}}-y_{ij}\sigma_{\alpha_{i}}^{2},\sigma_{\alpha_{i}}^{2}\right)\right],\\
q^{\prime}\left(u_{i}\right) & =\text{proj}\left[N\left(u_{i};\mu_{g},\Sigma_{g}\right)+c N\left(u_{i};\left(\Sigma_{u_{i}}^{-1}-\Sigma_{v_{j}}\right)^{-1}\left(\Sigma_{u_{i}}^{-1}\mu_{u_{i}}-y_{ij}\mu_{v_{j}}\right),\left(\Sigma_{u_{i}}^{-1}-\Sigma_{v_{j}}\right)^{-1}\right)\right],
\label{eq1:linear combination gaussian}
\end{align}
where
\begin{equation}
c=E_{\mu}\left[\exp\left(-y_{ij}\mu\right)\right]E_{\alpha_{i}}\left[\exp\left(-y_{ij}\alpha_{i}\right)\right]E_{\beta_{j}}\left[\exp\left(-y_{ij}\beta_{j}\right)\right]E_{u_{i},v_{j}}\left[\exp\left(-y_{ij}u_{i}^{T}v_{j}\right)\right]
\label{eq1:constant}
\end{equation}
and first and second moments can be calculated from each expression (after normalizing by $\frac{1}{1+c}$) to derive the corresponding Gaussian parameters.

To recap, our message passing algorithm will proceed as follows:
\singlespacing
\begin{enumerate}
\item Initialize all messages $m_{\left(i,j\right)\rightarrow\theta_{k}}$
\item Repeat until convergence of all messages:
\begin{enumerate}
\item Choose factor $\left(i,j\right)$
\item Update approximation to posterior $q\left(\theta\right)$ via equations (\ref{eq1:update step}) and (\ref{eq1:update step 2}). We find the choice of $\epsilon = 2$ promising in simulations.
\item Update messages from this factor $m_{\left(i,j\right)\rightarrow\theta_{k}}$ via equation (\ref{eq1:message update step})
\end{enumerate}
\end{enumerate}
\doublespacing

\subsection{Case-Control Approximation} \label{section1:case control}

The update step for each factor has $O(1)$ computational cost, but each iteration over the entire network has $O(N^2)$ computational cost and can be prohibitively expensive in large networks. In addition, tracking the messages for each factor also has $O(N^2)$ storage complexity. Drawing inspiration from \citep{raftery2012fast}, we wish to take advantage of the idea that large networks tend to be sparse and iterating over the entire network can be computationally inefficient due to the extreme class imbalance. The influence contained in each non-edge may be relatively small towards informing the overall model compared to the influence of edges, which are many fewer in number.

We propose iterating over the set of factors with $y_{i,j} = 1$ and a random sample of factors with $y_{i,j} = 0$. In practice, drawing a single random sample is preferable to drawing a new sample each iteration through the data due to the reduced time observed for the convergence of the algorithm. In a temporal setting, a new sample can be drawn at each time step. Supposing the number of observed edges $E = |\{i,j\}: y_{i,j} = 1| << N^2$ and a random sample of non-edges of size $O(E)$ is drawn, each iteration over the network would cost $O(E)$ rather than $O(N^2)$. Algorithmically, we treat this sample of factors as if it is the full set of data available. To understand the effect of this choice on the means of our parameter estimates, consider exponentiating both sides of equation (\ref{eq1:logit_bmem}):
\begin{equation}
    \frac{p_{i,j}}{1-p_{i,j}} = \text{exp}(\mu + \alpha_i + \beta_j + u_i^Tv_j).
\label{eq1:odds_ratio}
\end{equation}
Intuitively, sampling a random proportion $q$ of the non-edges inflates the odds-ratio on the LHS by a factor of $q^{-1}$. On the RHS, $\mu$ would shift upwards by $-\text{log}(q)$ but the other parameters would be unaffected. This suggests a simple post-hoc mean correction would suffice to return the parameters to their original scale, although it should be noted the posterior variance of the latent variables should be larger than if the full dataset were used. We provide some evidence for the efficacy of this case-control approximation in Section \ref{section1:simulation}.

\subsection{Temporal Dynamics} \label{section1:dynamics model}
Next, we introduce discrete time dynamics to the bilinear mixed-effects model by allowing each of the latent variables to evolve via a Markov chain. Let $\theta_{t,k}$ denote the $k$th parameter at time $t$, with the posterior of $\theta_{t,k}$ given (approximated) by
\begin{equation}
\theta_{t,k}|\textbf{Y}_{1:t} \sim N(\widehat{\mu}_{\theta_{t,k}}, \widehat{\Sigma}_{\theta_{t,k}}).
    \label{eq1:posterior}
\end{equation}
Supposing $\theta_{t,k}$ evolves via the Gaussian random walk
\begin{equation}
    \theta_{t+1,k} \sim \theta_{t,k} + N(0,W_{t+1,k}),
    \label{eq1:random walk}
\end{equation} the prior for $\theta_{t+1,k}$ would be given by
\begin{equation}
  \theta_{t+1,k}|\textbf{Y}_{1:t} \sim N(\widehat{\mu}_{\theta_{t,k}}, \widehat{\Sigma}_{\theta_{t,k}} + W_{t+1,k}).
\label{eq1:random walk prior fake}
\end{equation}

{Combining \ref{eq1:posterior} and \ref{eq1:random walk}, we could get the final form of theta posterior evolution over time as \ref{eq1:random walk prior fake}}.
We adopt the adaptive tuning procedure described in \citep{mccormick2012dynamic} to determine the amount of additional variation to introduce at each time point. We parametrize this amount of variation via a ``forgetting" multiplier $\tau_{t+1,k}\geq1$ to avoid specifying $d \times d$ random walk matrices for $u_i$ and $v_j$, {recasting (\ref{eq1:random walk prior fake}) as}:
\begin{equation}
  \theta_{t+1,k}|\textbf{Y}_{1:t} \sim N(\widehat{\mu}_{\theta_{t,k}}, \tau_{t+1,k}\widehat{\Sigma}_{\theta_{t,k}}).
\label{eq1:random walk prior actual}
\end{equation}
We choose these multipliers $\boldsymbol{\tau}_{t+1}$ based on the average predictive likelihood:
\begin{equation}
    \boldsymbol{\tau_{t+1}} = \underset{\boldsymbol{\tau}_{t+1}}{\text{argmax}}\frac{1}{N^2}\sum_{i,j}\int_{\theta_{t+1}}p(y_{t+1,i,j}|\theta_{t+1},\textbf{Y}_{1:t})p(\theta_{t+1}|\textbf{Y}_{1:t})d\theta_{t+1},
\label{eq1:average predictive likelihood}
\end{equation}
where $y_{t+1,ij}$ denotes the observation of $y_{ij}$ at time $t+1$. Evaluating the integral above cannot be done in closed form, and we use a series of two approximations to estimate it. First, note the likelihood term primarily involves the sigmoid of the logistic mean function:
\begin{align}
    p(1|\theta_{t+1},\textbf{Y}_{1:t}) &= p_{t+1} = \text{expit}\left(\mu_{t+1} + \alpha_{t+1,i} + \beta_{t+1,j} + u_{t+1,i}^Tv_{t+1,j}\right),\\
    p(0|\theta_{t+1},\textbf{Y}_{1:t}) &= 1-p_{t+1}.
\label{eq1:sigmoids}
\end{align}
Recall the prior over each latent variable is an independent Gaussian. We approximate $u_{t+1,i}^Tv_{t+1,j}$ with a single Gaussian term (via their first two moments) in order to model the entire mean function itself with a single Gaussian. Denoting the mean function with $\psi$, we use the following approximation for convoluting a sigmoid and a Gaussian could be used (see \citep{bishop2006pattern}):
\begin{equation}
\int \text{expit}(\psi)N(\psi |\mu_{\psi},\sigma^2_{\psi}) = \text{expit}((1+\pi\sigma^2_{\psi}/8)^{-1/2}\mu_{\psi}).
\label{eq1:convolution approx}
\end{equation}
Lastly, in order to reduce the computational cost involved in maximizing (\ref{eq1:average predictive likelihood}), we allow for a single forgetting multiplier for $\mu_{t+1}$, a single multiplier for the popularity terms $\alpha_{t+1,i}$ and $\beta_{t+1,j}$, and a single multiplier for the latent space terms $u_{t+1,i}$ and $v_{t+1,j}$, and only evaluate (\ref{eq1:average predictive likelihood}) over a coarse grid of values. \citet{mccormick2012dynamic} argue searching over a coarse grid leads to comparable results to directly maximizing (\ref{eq1:average predictive likelihood}) when running the algorithm over sufficiently many time periods, as periods with unnecessary inflation in prior variance can be balanced against periods with more restrictive inflation. Furthermore, the authors found their results were robust to the choice of grid values. In Sections \ref{section1:simulation} and \ref{section1:LANL data}, we take $\tau \in \{1,1.01,1.1,2\}$. This choice of values allows for no change in a parameter ($\tau = 0$), as well as forgetting multipliers corresponding to multiple scales of variance inflation.

\subsection{Anomaly Detection} \label{section1:anomaly detection}

At an edge level, anomaly detection proceeds by scoring edges based on their probabilities for observing activity, under the assumption that any past activity is non-anomalous and thus is a good representation of normal behavior. We score the dyad $i\rightarrow j$ at time $t$ via its predictive likelihood:
\begin{equation}
    \hat{p}_{i,j,t+1} \equiv \int_{\theta_{t+1}}p(y_{i,j}|\theta_{t+1},\textbf{Y}_{1:t})p(\theta_{t+1}|\textbf{Y}_{1:t})d\theta_{t+1},
    \label{eq1:predictive likelihood}
\end{equation}
which can be evaluated via the approximations described in the previous section. Dyads with activity but low predictive scores as well as dyads without activity but high predictive scores would then be flagged as anomalous.

For many settings, we may be interested in detecting anomalies at a non-edge level. For example, in computer networks such as the LANL network described in Section \ref{section1:Introduction}, security experts are interested in identifying anomalous subgraphs which may potentially represent intruder attacks. \citep{neil2013detection} mentions activity in the shape of $k$-stars and $k$-paths as common behavior for intrusions. Note dyads in these subgraphs would consist solely of edges with observed activity, so lower values of $\hat{p}_{i,j,t+1}$ would be characterized as more anomalous. We can compute scores for these subgraphs from our edge level scores given a conditional independence assumption, by multiplying the scores of the corresponding edges, or equivalently, summing the log scores.
\begin{figure}[ht]
    \centering
    \begin{tikzpicture}
    \node[draw, circle] (1) at (-4.5,0) {$1$};
    \node[draw, circle] (2) at (-3,0) {$2$};
    \node[draw, circle] (3) at (-1.5,0) {$3$};
    \node[draw, circle] (4) at (0,0) {$4$};

    \path[->] (1) edge node {} (2);
    \path[->] (2) edge node {} (3);
    \path[->] (3) edge node {} (4);
    
    \end{tikzpicture}
    \caption{A directed 3-path.}
    \label{fig1:tikz subgraphs}
\end{figure}
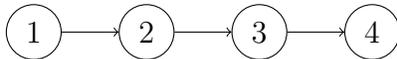
For example, for the 3-path shown in figure \ref{fig1:tikz subgraphs}, the score would be given by $\hat{p}_{1,2,t+1}\hat{p}_{2,3,t+1}\hat{p}_{3,4,t+1}$. 

In this paper, we choose to separately consider potentially anomalous behavior for each time period, although combining scores across time may be promising, particularly in settings with fine temporal resolution where attacks may span multiple periods. A fully online detection procedure would proceed at each time step as follows:
\begin{enumerate}
    \singlespacing
    \item Observe network behavior $\mathbf{Y}_t$
    \item Tune forgetting multipliers (\ref{eq1:average predictive likelihood})
    \item Flag and assess potential anomalous subgraphs
    \item Remove anomalous activity from $\mathbf{Y}_t$
    \item Estimate model parameters $\Theta_t$ 
\end{enumerate}
\doublespacing

\section{Simulation Studies} \label{section1:simulation}
{In this section we provide results from two simulation experiments to evaluate the performance of our proposed approach.  First, we conduct a simulation where network data are generated using the bilinear mixed-effects model, in accordance with our model assumptions.  We also perform a second simulation when data arise from a different network generating process.  Additional results that demonstrate the robustness of our work to tuning parameter choices is presented in the supplementary appendix.}

First, To provide an idea of how well our proposed algorithm can estimate a dynamic bilinear mixed-effects model, we simulate a network following  equations (\ref{eq1:logit_eq}) and (\ref{eq1:logit_bmem}) and with time dynamics following independent Gaussian random walks (see (\ref{eq1:random walk})). Specifically, we generate a network of size $N=500$ from a  bilinear mixed-effects model with latent dimension 2 with the following priors:

\begin{align}
    \mu_{1} & \sim N(-6.5,0.1),\\
    \alpha_{1,i} & \sim N(0,1),\\
    \beta_{1,j} & \sim N(0,1),\\
    u_{1,i} & \sim N\left(\begin{pmatrix} 0 \\ 0 \end{pmatrix},\begin{bmatrix} 0.75 & 0.15 \\ 0.15 & 0.75 \end{bmatrix} \right),\\
    v_{1,j} & \sim N\left(\begin{pmatrix} 0 \\ 0 \end{pmatrix},\begin{bmatrix} 0.75 & 0.15 \\ 0.15 & 0.75 \end{bmatrix} \right).
    \label{eq1:simulation priors}
\end{align}

We evolve the network 99 times for a total of $T = 100$ periods, where at every time point each parameter follows a Gaussian random walk with (co)variance equal to 0.001 times its prior (co)variance. These prior and random walk values were chosen to create a network that would be roughly similar to the LANL computer network, that is, characterized by high sparsity, strong heterogeneity between nodes, low temporal dynamics, and strong dependence between time periods. The generated network averages about 2,000 directed connections per time period, or 4 per node, which is slightly less than what we observe for the LANL network.

We compare results from two runs of the Power EP algorithm described in Section \ref{section1:vi}, one that iterates over all 250,000 potential dyads in the network and another that implements the case-control modification described in Section \ref{section1:case control}. For the latter, we sample 2.5\% of the non-edges at every time point for consideration, thus iterating over about 8,200 dyads per time point. This results in about a 93.5\% reduction in computation time and a 97\% reduction in storage complexity.

\begin{figure}[ht]
    \centering
    \includegraphics[width=.8\linewidth]{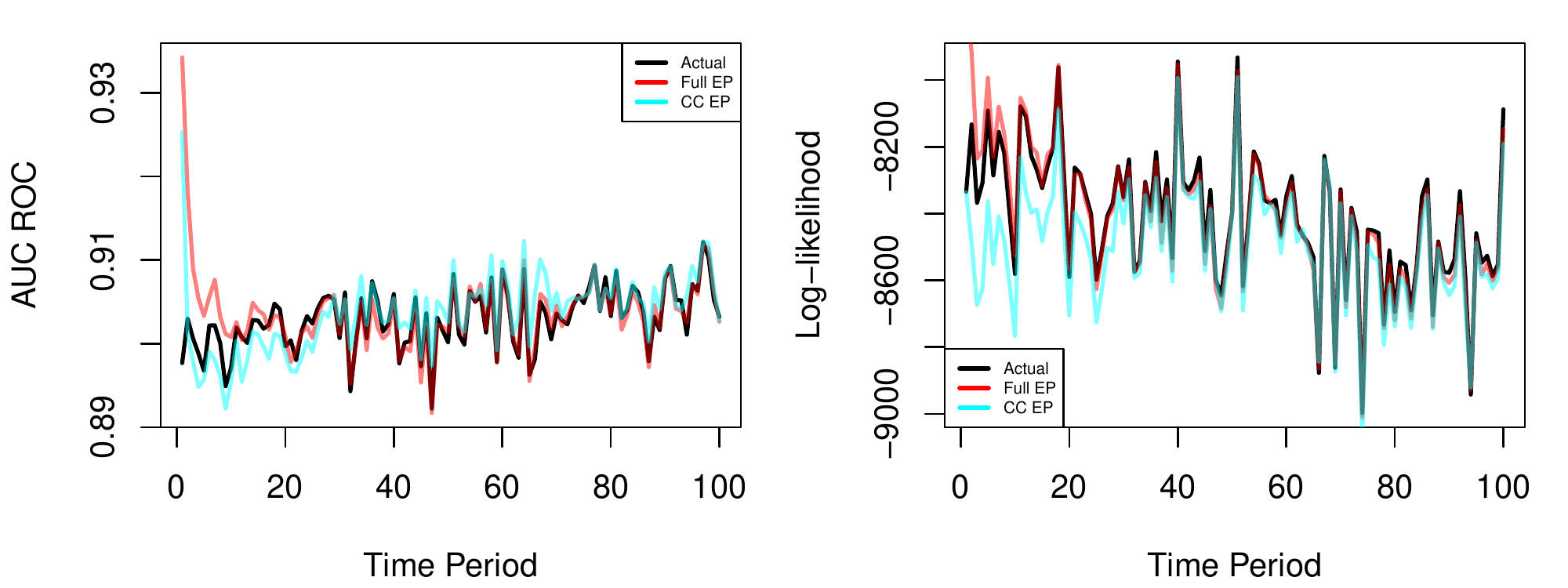}
    \caption{Log-likelihood of the observed network $\mathbf{Y}_t$ and ROC AUC of the edge probabilities $\hat{p}_{i,j,t}$. Calculated for three sets of edge probabilities: the actual edge probabilities (black), the edge probabilities estimated via the Power EP algorithm on the full data (red), and the edge probabilities estimated via the Power EP algorithm with the case-control approximation (cyan).}
    \label{fig1:simulation_ll_auc}
\end{figure}

We compare mean estimates from each run against the generated values in terms of log-likelihood, area under the receiver operating characteristic curve (ROC AUC), and the correlation between the actual and estimated edge probabilities on the logit scale. This correlation is also calculated restricted to dyads not observed in any of the 100 time periods (which is satisfied by 72.6\% of all dyads). In Figure \ref{fig1:simulation_ll_auc}, we compare the model fit of the mean estimates from the Power EP algorithm with and without the case-control approximation to the model fit of the true parameter values. In these plots, the log-likelihood and ROC AUC under the generating model provide a soft bound on model performance, as no other set of parameter estimates should systematically outperform them over a prolonged period. The estimates from our variational approach perform very similarly to the generating model, particularly after time period 40, suggesting the approximations used in the variational method have, at most, minor effects on the mean posterior estimates. 

\begin{figure}[ht]
    \centering
    \includegraphics[width=.8\linewidth]{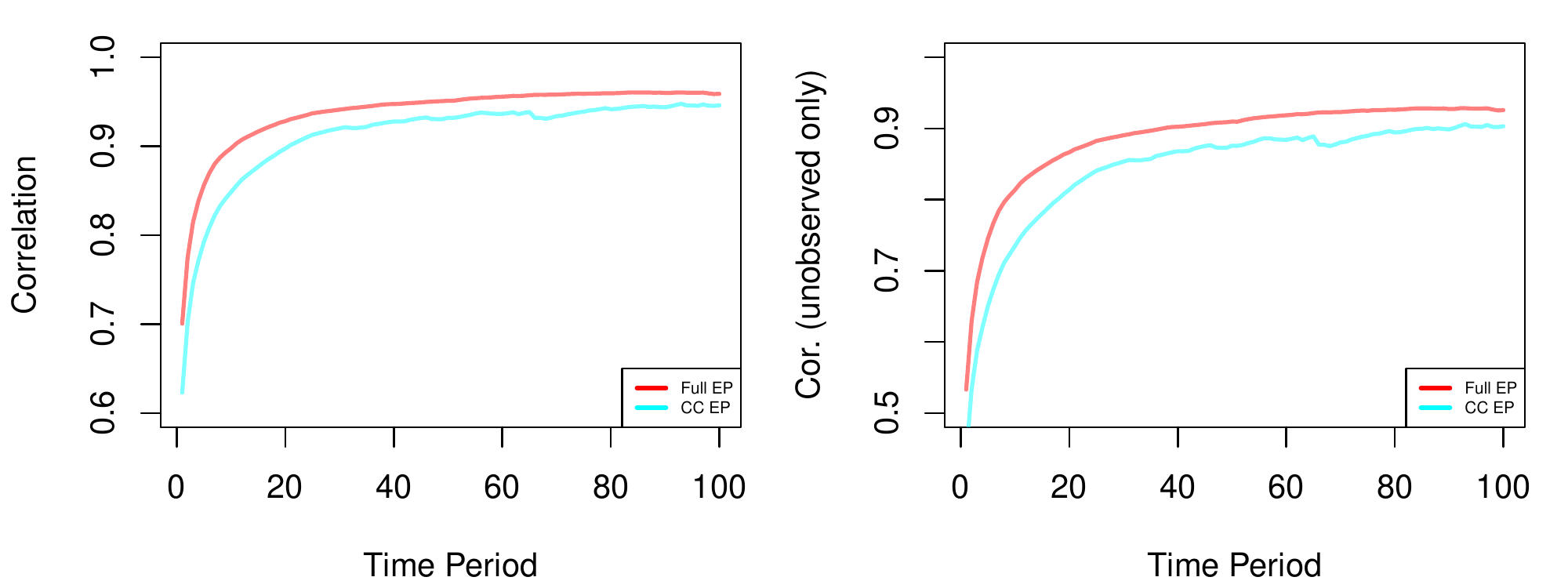}
    \caption{Correlation between the actual edge probabilities and estimated edge probabilities, both on the logit scale. Results with probabilities from the full data are in red, and results with probabilities from the case-control are in cyan.}
    \label{fig1:simulation_corr}
\end{figure}

In Figure \ref{fig1:simulation_corr}, we plot the correlation between the actual edge probabilities and their estimated counterparts on the logit scale, using both the full variational approach and the case-control approximation. The performance of the algorithms ramps up over time, as each binary network $\mathbf{Y}$ provides limited information about the underlying latent variables which must be aggregated, and is largely stabilized by time 40. The case-control modified algorithm, which iterates over a much smaller subset of the network at each time point, does perform worse than the algorithm over the full network, but these differences are largest in the earlier time periods.

\begin{figure}[ht]
    \centering
    \includegraphics[width=.65\linewidth]{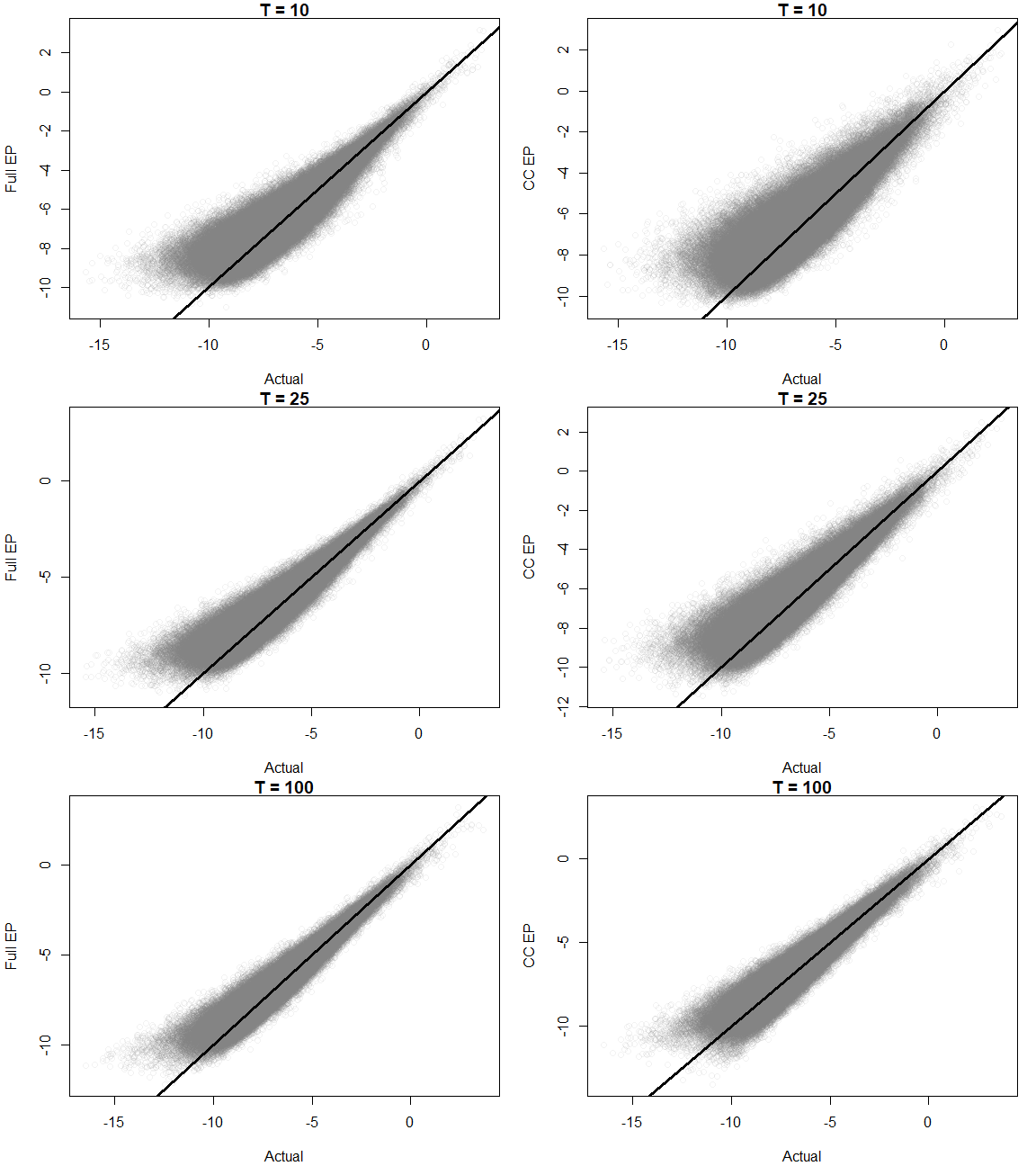}
    \caption{Actual versus estimated edge probabilities on the logit scale. The left column compares the actual edge probabilities to estimates from the full Power EP, while the right column represents the edge probabilities estimated via the case-control Power EP.}
    \label{fig1:scatter}
\end{figure}

 Restricting ourselves to results from three time points, we plot the distributions of the edge probabilities (again on the logit scale) against their actual counterparts in Figure \ref{fig1:scatter}, and find minor systematic differences between the distributions. Note both sets of estimated probabilities do struggle a bit (overestimating) modeling the extreme left tail of probabilities, although these differences are exacerbated due to the logit scale (e.g. $\text{expit}(-14)$ = 8.3e-07 and $\text{expit}(-16)$ = 1.1e-07) and may be hard to capture given the time frame of the simulation in comparison to the probability size.  {We show a similar plot for the model parameters in the Appendix in Figure~\ref{fig1:alpha}.  Table~\ref{table1:posterior var} describes the increase in posterior variance of our parameters when adopting the case-control modification to be largely acceptable.  Even though we only consider about 3\% of the edges in any given time period, this subset of the network appears to capture most of the information for estimating the model parameters. We demonstrate with Figure \ref{fig1:simulation_time} in the appendix that the case-control approximation had roughly a ten-fold decrease in computation time at each update compared to the full EP approach, plus a substantial reduction in estimating parameters in the first time period.}
 
\begin{table}[ht]
\caption{Multiplier in posterior variance when using the case-control modification to the Power EP algorithm. For node-specific parameters, the multiplier in variance is calculated for each node and averaged.}
\label{table1:posterior var}
\begin{tabular}{l|lllll}
Time Period & $\mu$ & $\alpha_i$ & $\beta_j$ & $u_{i,1}$ & $v_{j,2}$ \\
\hline
T = 10 & 6.76 & 1.92 & 1.97 & 2.11 & 2.42 \\
T = 25 & 1.92 & 2.47 & 2.55 & 2.10 & 2.32 \\
T = 100 & 2.24 & 2.63 & 2.73 & 1.54 & 1.78 
\end{tabular}
\end{table}

{We now perform a second simulation exercise where the underlying model differs from the model we use in our approach.  Based on the work of \citet{zhao2018temporal}, we generated a degree corrected
stochastic block model (DCSBM). It is briefly described below and the details of the DCSBM network model can be found in \citet{zhao2018temporal}.} In DCSBM model, all the nodes are separated into R communities, denoted by $r=1,2,...,R$. The number of edges $C_t(i,j)$ formed between node $i$ and node $j$ follows a Poisson distribution
$C_t(i,j)\sim Poisson(\lambda_{i,j})$.
Where $\lambda_{i,j}$ is calculated as $\lambda_{i,j}=\theta_i\theta_j P_{r,r'}$ and 
$P_{r,r'}$ is the communication propensity between communities $r$ and $r'$, which measures how likely a member in community $r$ connect with a member in community $r'$. $\theta_i$ is obtained by the following procedure. First, we have a series of parameters $\theta^{'}_{i}$ which follow identically independently distributed Pareto Distribution.

\[\theta'_i\sim PAR(m=1, s=3)\]
Then $\theta_i$ is calculated by 
\[\theta_i=\frac{\theta'_i}{\sum_{i=1}^{|V_r|\theta'_i}}\times|V_r|\]
where $|V_r|$ is the number of members in community $r$. 

\begin{figure}
    \centering
    \includegraphics[width=.8\textwidth]{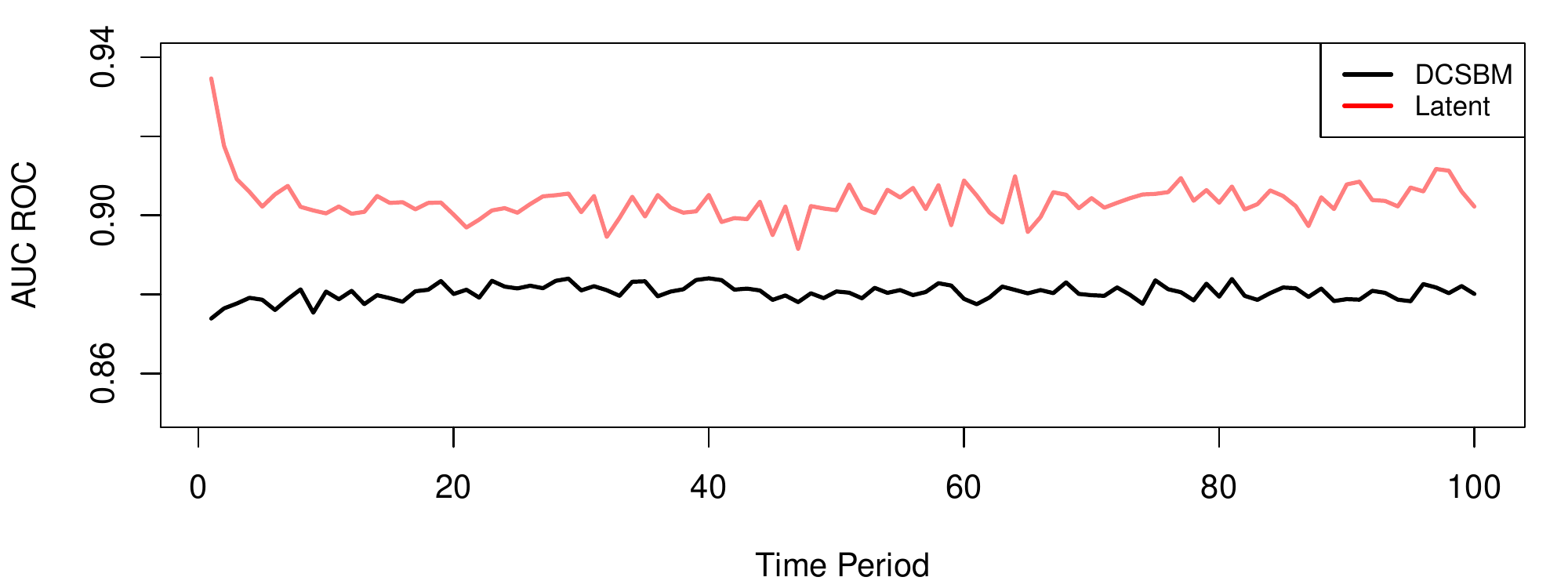}
    \caption{Area under the ROC curve for models generated by two network models.  The red line represents the bilinear mixed effects model and the black is the degree-corrected stochastic blockmodel.}
    \label{fig:dcsbm}
\end{figure}

{
In our simulation study, we generated a DCSBM network of 500 nodes and randomly separated the nodes into 20 communities. A total of $T = 100$ periods were simulated. For the first 50 periods, The within community propensity is $P_{r,r'}=0.2$ if $r=r'$ and the between community propensity is $P_{r,r'}=0.1$ if $r \neq r'$. Starting from 51st period, the within community propensity for the first community shifted to 0.5 while the between community propensity remains the same. This means that the members in the first community tend to communicate more within the community. We performed our method on the DCSBM network model and the results are shown as Figure~\ref{fig:dcsbm}. The AUC-ROC curve showed that our method could still reach close to 0.9 ROC values on the DCSBM network.}

\section{LANL Netflow Event Data} \label{section1:LANL data}

We demonstrate the potential for attack detection with Netflow communications data on the LANL enterpise network \citep{turcotte2018unified}. Event records correspond to directed communication between two network devices and span a total of 89 days. We restrict to the sub-network of the $N = 27,436$ computers with some record of outgoing communications over the 89 days, and aggregate the event records into four-hour intervals, yielding a total of $T=532$ time periods. These computers comprise the set of network devices which may be the source of malicious behavior. We focus on modeling the presence of any directed network activity between each dyad within each four-hour interval. The resulting network averages about 150,000 directed edges (or 5.5 outgoing edges for each computer) at each time interval, and there is substantial variation in activity levels based on time-of-day and day-of-week.

\begin{figure}[ht]
    \centering
    \begin{tikzpicture}[thick,scale=0.6, every node/.style={scale=0.6}]
    \node[draw, circle] (1) at (-4.5,0) {$1$};
    \node[draw, circle] (2) at (-3,0) {$2$};
    \node[draw, circle] (3) at (-1.5,0) {$3$};
    \node[draw, circle] (4) at (0,0) {$4$};
    
    \node[draw, circle] (9) at (2,0) {$1$};
    \node[draw, circle] (10) at (3.5,0) {$2$};
    \node[draw, circle] (11) at (4.56066,-1.06066) {$3$};
    \node[draw, circle] (12) at (4.56066,1.06066) {$4$};
    
    \node[draw, circle] (5) at (6.5,0) {$1$};
    \node[draw, circle] (6) at (7.56066,1.06066) {$2$};
    \node[draw, circle] (7) at (8,0) {$3$};
    \node[draw, circle] (8) at (7.56066,-1.06066) {$4$};

    \path[->] (1) edge node {} (2);
    \path[->] (2) edge node {} (3);
    \path[->] (3) edge node {} (4);

    \path[->] (9) edge node {} (10);
    \path[->] (10) edge node {} (11);
    \path[->] (10) edge node {} (12);

    \path[->] (5) edge node {} (6);
    \path[->] (5) edge node {} (7);
    \path[->] (5) edge node {} (8);
    \end{tikzpicture}
        \caption{A 3-path (left), a 3-star (right), and a ``fork" (center) representing a combination of the two.}
    \label{fig1:more tikz subgraphs}

\end{figure}
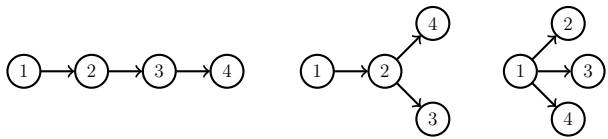

The LANL data contains a red team attack in the form of a network scanning attack from ``Computer A" that begins on day 57, and we are interested in the ability for our model to recognize this activity as anomalous. Following \citet{neil2013detection}, we detect potentially anomalous subgraphs of the three shapes presented in Figure \ref{fig1:more tikz subgraphs}, corresponding to common intrusion patterns. While malicious attacks may involve more nodes and activity, detecting a single subgraph involved in the attack may suffice to identify the entire attack upon further (manual) examination. 
Note the detection procedure described in this section deviates from the typical online setting since details of the red team attack were only obtained after model estimation. Flagging potential anomalies only occurs after model estimation, and the estimated probabilities used in this process assume all preceding network activity was non-anomalous.

We estimate the bilinear mixed-effects model with latent dimension $d=2$ using the Power EP approach described in Section \ref{section1:vi} with the case-control approximation of Section \ref{section1:case control}, taking a sample of the non-edges of average size 500,000 (corresponding to a case-control rate of 3.3 or sampling proportion $q\approx0.066\%$). We slightly modify the bilinear mixed-effects model to incorporate time-of-day and day-of-week effects in the form of mean shifts, with individual terms for each time-of-day and day-of-week pair calculated directly from the mean activity levels over the 89 days. A slightly more sophisticated approach would be to include them as additional parameters in the model to estimate. This would allow these effects to naturally change over time, although there is little evidence for any such changes in the observed data. We choose to separately model these terms from the overall popularity term $\mu_t$ in order to prevent the dynamics of this parameter to be governed by periodicity effects rather than random walk behavior \citep{heard2014filtering}. Note that part of the periodicity effects may be due to recurrent, automated tasks (e.g. weekly at a certain time of day), so allowing for more complicated periodicity effects or removing these activities before estimation (if they are labeled or can be \textit{a priori} identified) would likely improve model fit.

\begin{figure}
    \centering
    \includegraphics{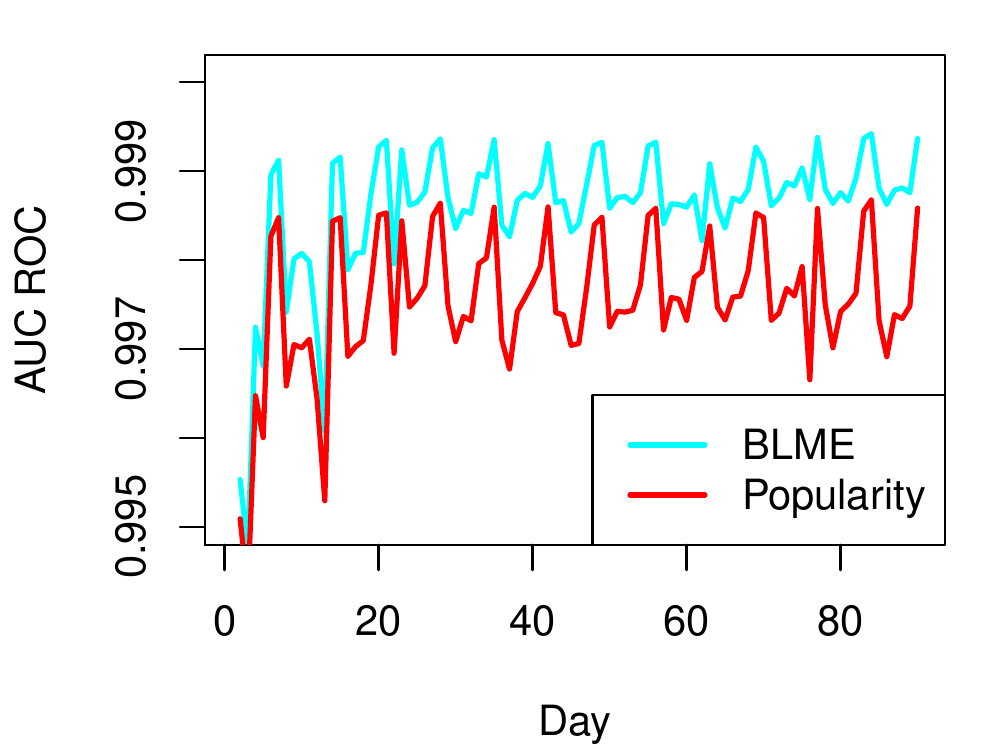}
    \caption{Averaged AUC ROC for popularity model (red) and bilinear mixed-effects model (cyan) over days. AUC ROC is calculated for each 4-hour interval using probabilities derived from the predictive likelihood and results are averaged across each day.}
    \label{fig1:auc_lanl}
\end{figure}

Before turning to anomaly detection, we assess how well the bilinear mixed-effects model and the popularity model omitting the latent interaction terms are able to predict LANL network activity. Figure \ref{fig1:auc_lanl} plots the area under the receiver operating characteristic curve (AUC ROC) of both models calculated using probabilities from the predictive likelihood (\ref{eq1:predictive likelihood}), which are primarily dependent on estimated parameters from the previous time period. Both models perform quite well, with AUC $>$ 0.995, suggesting the network communications data is inherently very structured and predictable in nature. Model performance exhibits both time-of-day and day-of-week periodicity, suggesting a more complex approach to modeling periodicity is likely to improve performance, albeit performance is consistently high despite these effects. The latent interaction terms in the bilinear mixed-effects model do seem to substantially improve performance, with these improvements primarily driven by higher probabilities for active dyads with generally low overall levels of popularity in the network.

\begin{figure}
    \centering
\includegraphics[scale=0.6]{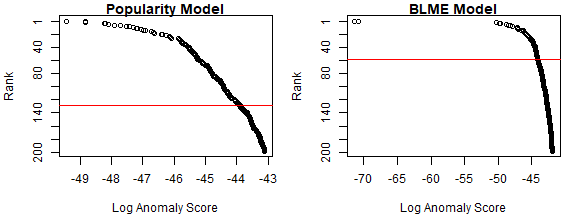}
    \caption{Anomaly scores for the 200 lowest scoring subgraphs observed over the 89 day period. The red line corresponds to the rank of a subgraph containing part of a red team attack on day 57.}
    \label{fig1:anomaly_score_dist}
\end{figure}

We can compute anomaly scores for subgraphs of the types shown in Figure \ref{fig1:more tikz subgraphs} by taking the sum of the log probabilities as described in Section \ref{section1:anomaly detection} . Despite the relative sparsity of the network, the number of subgraphs to consider at each time frame remains quite large. To reduce the number of subgraphs in consideration further, we only examine subgraphs consisting of edges with log probability score of -10 or lower and remove some ``overlapping" subgraphs. Specifically, we remove 3-paths with the same middle edge $``2"\rightarrow ``3"$, forks with the same $``2"$ node, and 3-stars with the same $``1"$ node. Ideally, detecting one anomaly from multiple overlapping subgraphs would suffice for finding the entire attack. In Figure \ref{fig1:anomaly_score_dist}, we plot the 200 most anomalous subgraphs under each model. Both sets of subgraphs contain a single 3-star (highlighted in red) involving the network scanning attack from Computer A on the first day of the red team attack. The rank of the Computer A 3-star is twice as high under the bilinear mixed-effects model, where this anomaly would be detectable given an average alarm rate of one subgraph per day. The difference in rank can be mainly attributed to low scores on recurrent activity between low-popularity computers under the popularity model, which is not flexible enough to model such activity. This leads to lower scores for certain non-anomalous subgraphs, obfuscating the actual attack. Note our detection procedure is largely unable to identify other attacks from Computer A in the following days. Once anomalous behavior like the activity on day 57 is incorporated into the model of normal activity, subsequent attacks appear to be normal behavior.

\begin{figure}[ht]
    \centering
    \includegraphics[width=.8\linewidth]{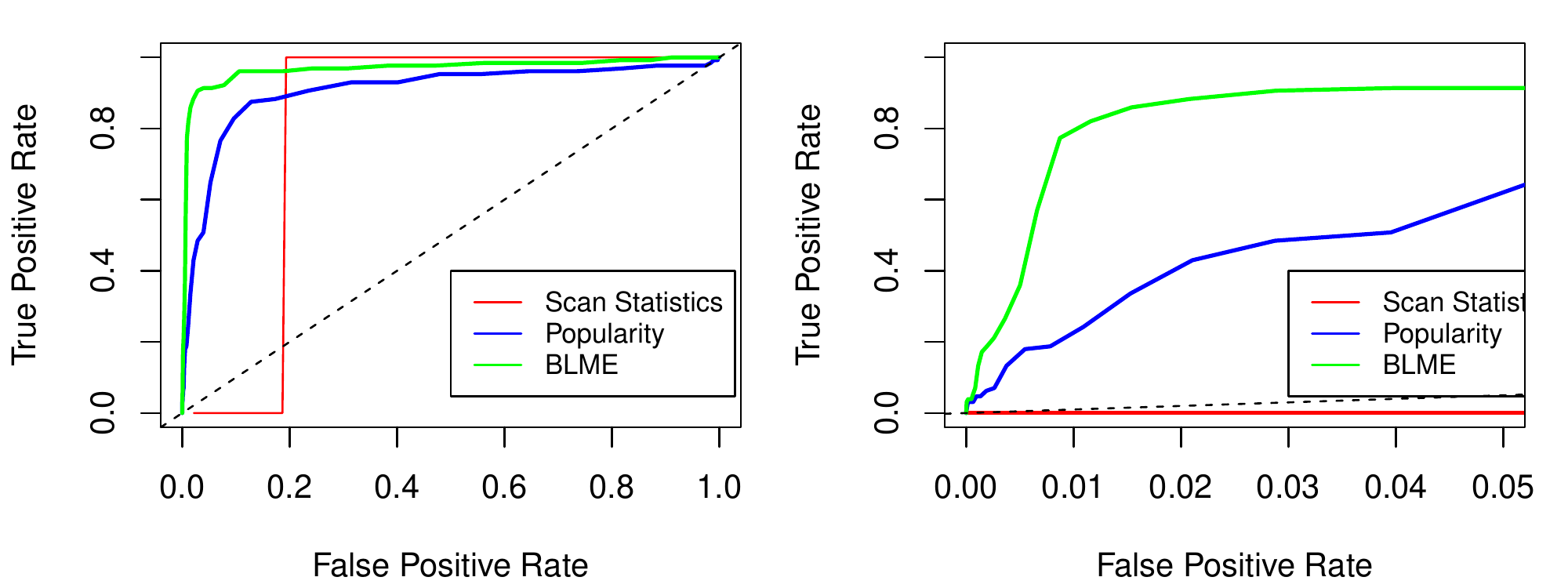}
    \caption{ROC curve of the LANL data at day 57. Three different methods are performed: scan statistics (red), Popularity model (blue) and BLME model (green).Left: Full-scale, Right: Zoom-in within 5\% false positive rate.}
    \label{fig1:lanl_tpr_fpr}
\end{figure}

{On day 57, the detection true positive rates and false positive rates are shown on the left of Figure \ref{fig1:lanl_tpr_fpr}. For comparison, we adopted the traditional scan statistics method for the anomaly detection as well \citet{priebe2005enron}. The scan statistics method is based on the number of edges of the subgraphs for each node center. Briefly, for each node, we calculate the neighbor nodes that can be connected to it with no more than 2 intermediate nodes. The number of edges of the subgraph formed by the neighbor nodes is calculated as scan statistics. The nodes with significant higher numbers of edges than their corresponding average number of edges for the same node in a certain past time window are considered as anomalies. However, this method is more node-oriented. In our application, the edges should be considered separately. For simplicity, when applying scan statistics method, all the edges connected with the anomaly node are considered as anomaly edges.

We also compared our method to one cased on scan statistics~\citep{priebe2005enron}. The scan statistics is basically counting the number of edges for a node and the nodes that have connections with it and take the average of the numbers within a time window. However, this method is less sensitive to the dynamical changes of the network. Also it is more efficient in detecting anomaly nodes instead of individual edges. The anomaly attacks mostly comes from a single computer at day 57, thus the scan statistics showed a sharp change in the true positive rate (either can't detect the anomaly node or can detect the single anomaly node, thus detecting all edges coming from it).}

\begin{figure}[ht]
    \centering
    \includegraphics[width=.8\linewidth]{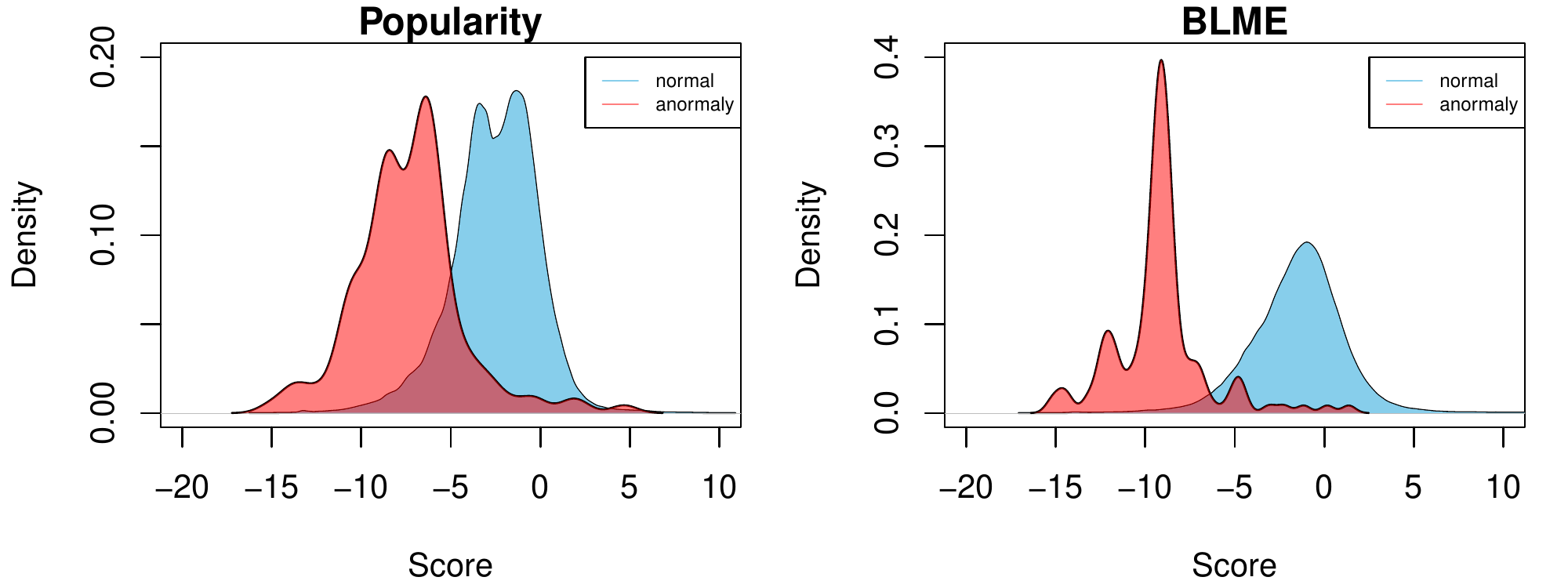}
    \caption{Score distribution for the normal and anomalous edges obtained by Popularity model (left) and BLME model (right). }
    \label{fig1:score_dist_lanl}
\end{figure}

{In applied cyber security use cases, false positive rates need to be kept extremely low, given high volumes of events flowing into the detectors, and the need to severely limit total detections for the human response teams that consume and react to these alerts. As a result, we are typically only interested in false positive rates in the 1\% range, where our algorithm clearly outperforms the baseline as shown by the right in Figure  \ref{fig1:lanl_tpr_fpr}. What’s more, node based anomalies can be scanning malware. This is a type of behavior of the attack in which the infected node reaches out through the network to locate servers that can be used to further the attack.  Other types of attack behavior such as lateral movement are composed of elongated attack structures known as ``caterpillars" \citet{neil2013detection}. In this setting, edge-based anomaly scoring is important to capture non-node centric attacks. Thus our method showed better flexibility in application for anomaly edge detection. 
The score distribution at day 57 for normal and anomalous edges produced by Popularity model and BLME model are shown in Figure \ref{fig1:score_dist_lanl}. As can be seen, both showed quite separable distributions between normal and anomalous edges and BLME model has better separation than Popularity model. This further demonstrated the efficiency of the anomaly detection method proposed here and explains the reason for good performance of the ROC curve.

}

\section{Discussion} \label{section1:Discussion}

In this paper, we present a variational approach for estimating the bilinear mixed-effects model. We adapt our approach to a dynamic, large network setting via a case-control approximation and an autotuning procedure to mimic Gaussian random walks on the model parameters. We demonstrated the efficacy of our algorithm via a simulation study on $N=500$ nodes, estimated the mixed-effects model on the LANL netflow communications network involving over 25,000 computers for a period of 89 days, and detected a red team attack on the same network while only requiring half the detection rate of the popularity model.

A natural extension for the bilinear mixed-effects model considered would be to allow for node- or edge-level covariates in the specification of the mean function. Relational event data is often provided with additional details which may be useful in conjunction with network-based predictors for predicting activity. Along a similar line of reasoning, edge-level covariate data may distinguish between multiple types of network activity which we may wish to model jointly. In particular, the LANL netflow data includes sender port information, and utilizing this information would help distinguish between typical activity on a commonly used port and unusual activity on a rarely used port. {Additionally, updating how the algorithm adjusts to new signals after detecting an anomaly could prevent anomalous behavior being integrated into the baseline.} Lastly, adapting the algorithm to handle different 
outcome measures may allow for a more faithful representation of the observed event data. For example, acknowledging the data's continuous-time nature, we could model the communication activity using Poisson processes with time dependent intensities in order to exploit detailed information about the timing of expected activity.

\bibliographystyle{abbrevnamed}
\bibliography{microsoft}

\pagebreak
\begin{appendix}
\centerline{\huge \bf Supplementary Appendix}
\section{Robustness} \label{section1:Robust Analysis}
\renewcommand\thefigure{\thesection.\arabic{figure}}  
\setcounter{figure}{0} 
{
In this section we present several robustness checks for our proposed method.  First, the damping factor $\epsilon$ and the sparsity of the network generated are varied to check the influences on the detection results. The detection results with our method for different settings are shown as Figure \ref{fig1:robust_analysis}. The first row of Figure \ref{fig1:robust_analysis} shows the detection results of different damping factors. For different damping factors, the AUC-ROC changes are minor as time periods progress, despite the initial difference. The correlation between the true model parameters and the estimated parameters all reach around 0.9 in the long run, so the damping factor only has minor impact on the final results and it is mostly used for convergence in the message passing algorithm.  The second row of Figure \ref{fig1:robust_analysis} shows the detection results of different sparsity of the simulated networks. The sparsity is denoted by the mean density of the edges, and ranges from 0.06\% to 1.0\%. At high sparsity (0.06\%), it takes more time periods to see high values of correlation and AUC-ROC. For the high sparsity network, we have fewer messages to update the parameters at each time point. As the sparsity decreases, our method performs well at earlier time periods.} 
\begin{figure}[ht]
    \centering
    \includegraphics[width=.8\linewidth]{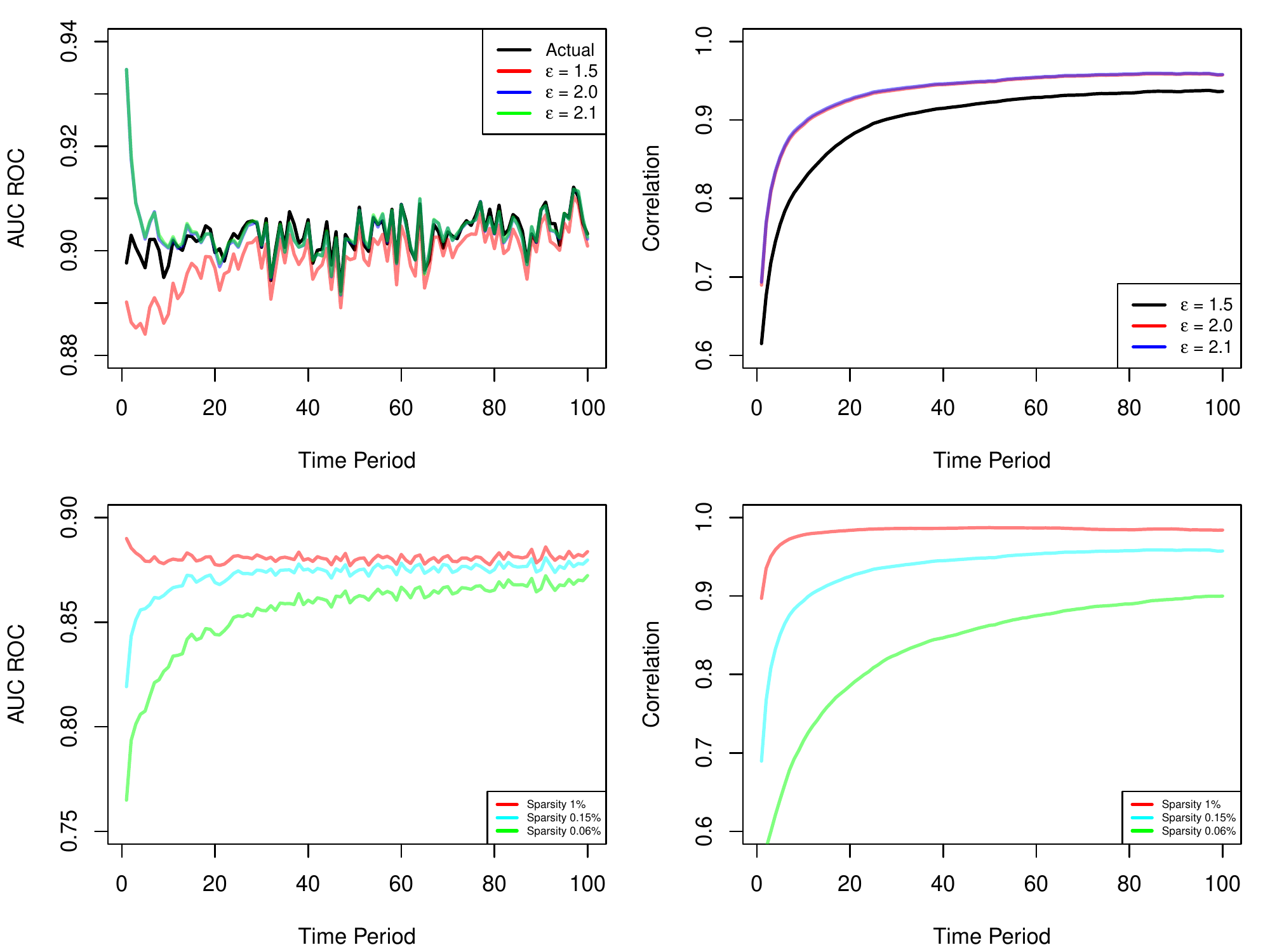}
    \caption{AUC-ROC values and the correlation between actual edge probabilities and estimated edge probabilities at different time periods.The first row is the results for different \(\epsilon\)values. The second row is the results for different sparsity.}
    \label{fig1:robust_analysis}
\end{figure}

\begin{figure}[ht]
    \centering
    \includegraphics[width=.8\linewidth]{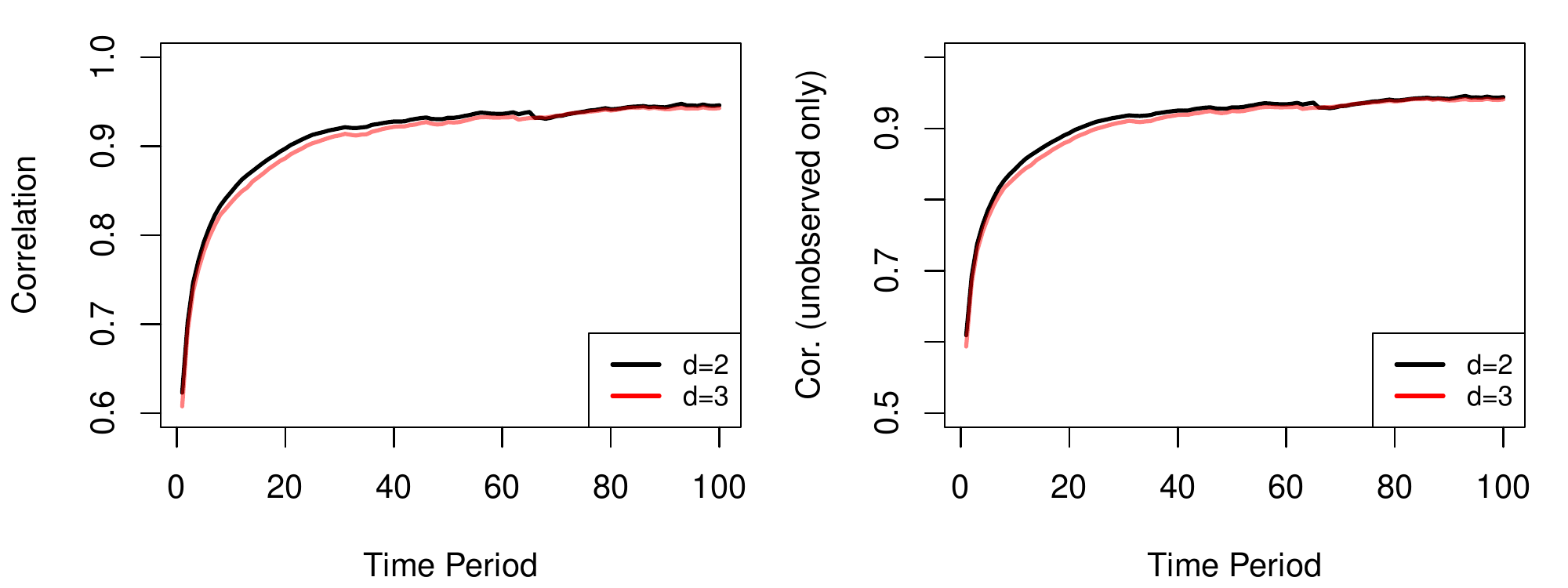}
    \caption{Correlation between the estimated edge probabilities and the actual counterparts on the logit-scale by different latent space dimensions.  The left panel shows all pairs while the right panel shows only unobserved edges.}
    \label{fig1:latent d3}
\end{figure}

{Next, we checked the influence of the dimension of the latent space. Although 2-dimensional latent space is utilized for our anomaly detection, 3-dimensional latent space is also checked and the detection results on the simulation data are shown as Figure \ref{fig1:latent d3}. As can be seen, increasing the dimension of the latent space doesn't bring significant difference in detection results. However, the computation cost is dramatically increased due to solving of large number of inverse matrices.}

\begin{figure}[ht]
    \centering
    \includegraphics[width=.5\linewidth]{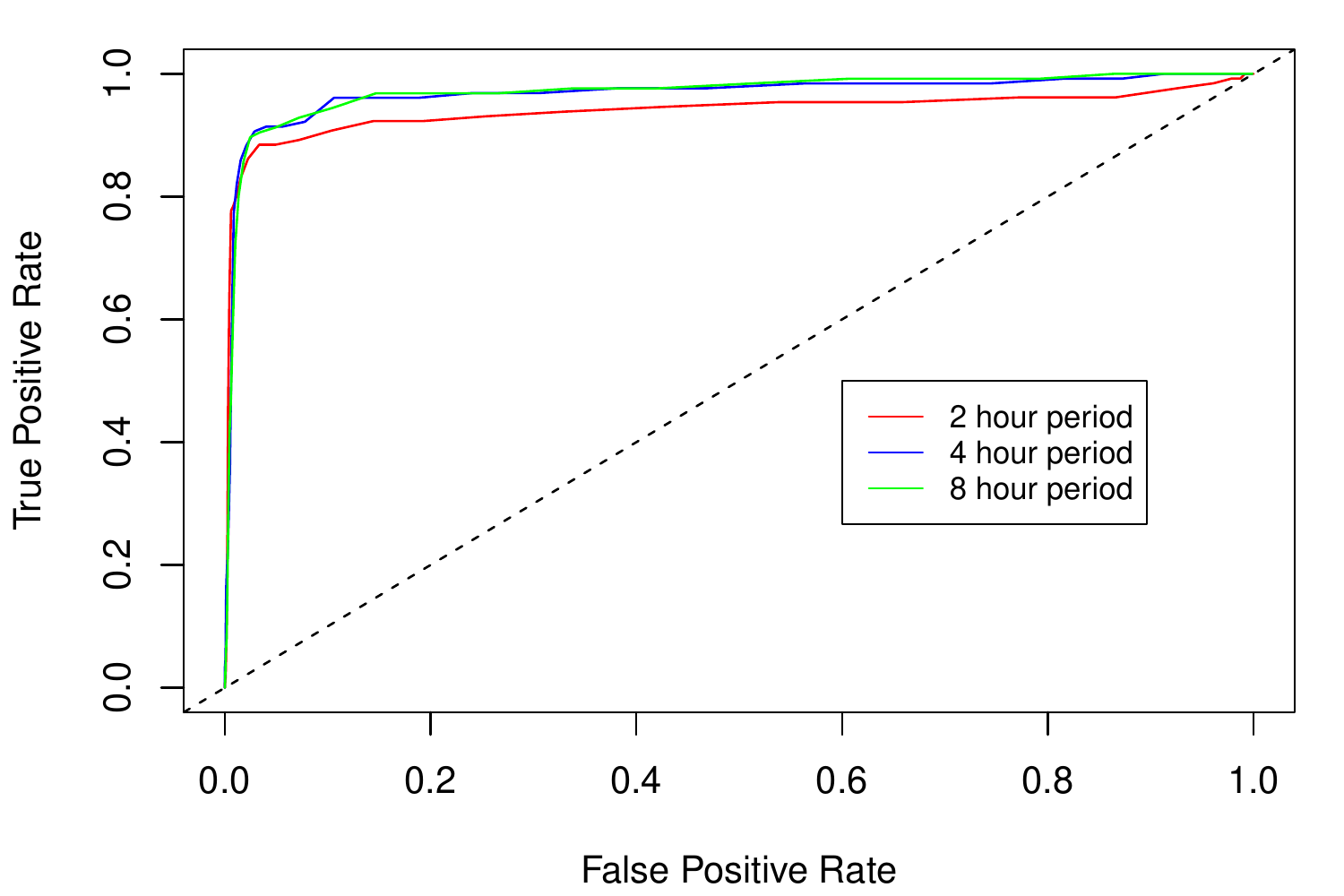}
    \caption{The ROC curve of the BLME model on the LANL data at day 57 with different time period separation: 2 hour/period, 4 hour/period and 8 hour/period. }
    \label{fig1:score_dist_lanl_time}
\end{figure}

{
Finally, we tried dividing records into different time periods for the LANL data. Figure~\ref{fig1:score_dist_lanl_time} shows the results using 2 hours/period, 4 hours/period and 8 hours/period. The 8 hours/period showed very similar detection results with 4 hours/period. The 2 hours showed slightly lower true positive rate, though the difference is again very small. Fewer messages are available in each 2 hour period, so we expect the parameter estimate updates have less information at each time period. However, all the time periods considered showed high AUC-ROC values ($>$ 0.9) for anomaly detection.}

\section{Evaluating the case-control approximation}
\setcounter{figure}{0} 
{In this section we present additional results comparing the case-control approximation to the full method.  In order to compare the estimated parameters and the true counterparts, we plotted the estimated $\alpha_i$ values for each node at different time periods. This is shown in Figure \ref{fig1:alpha}. Both full EP method and case-control EP method showed very similar estimation results and the case-control EP method showed little difference compared to the full EP method. Additionally, the computation time is shown in Figure \ref{fig1:simulation_time}. The case-control method significantly reduced the computation cost for the anomaly detection, particularly in the first period during parameter estimation.  This advantage persists, however, throughout subsequent periods.}

\begin{figure}[ht]
    \centering
    \includegraphics[width=.6\linewidth]{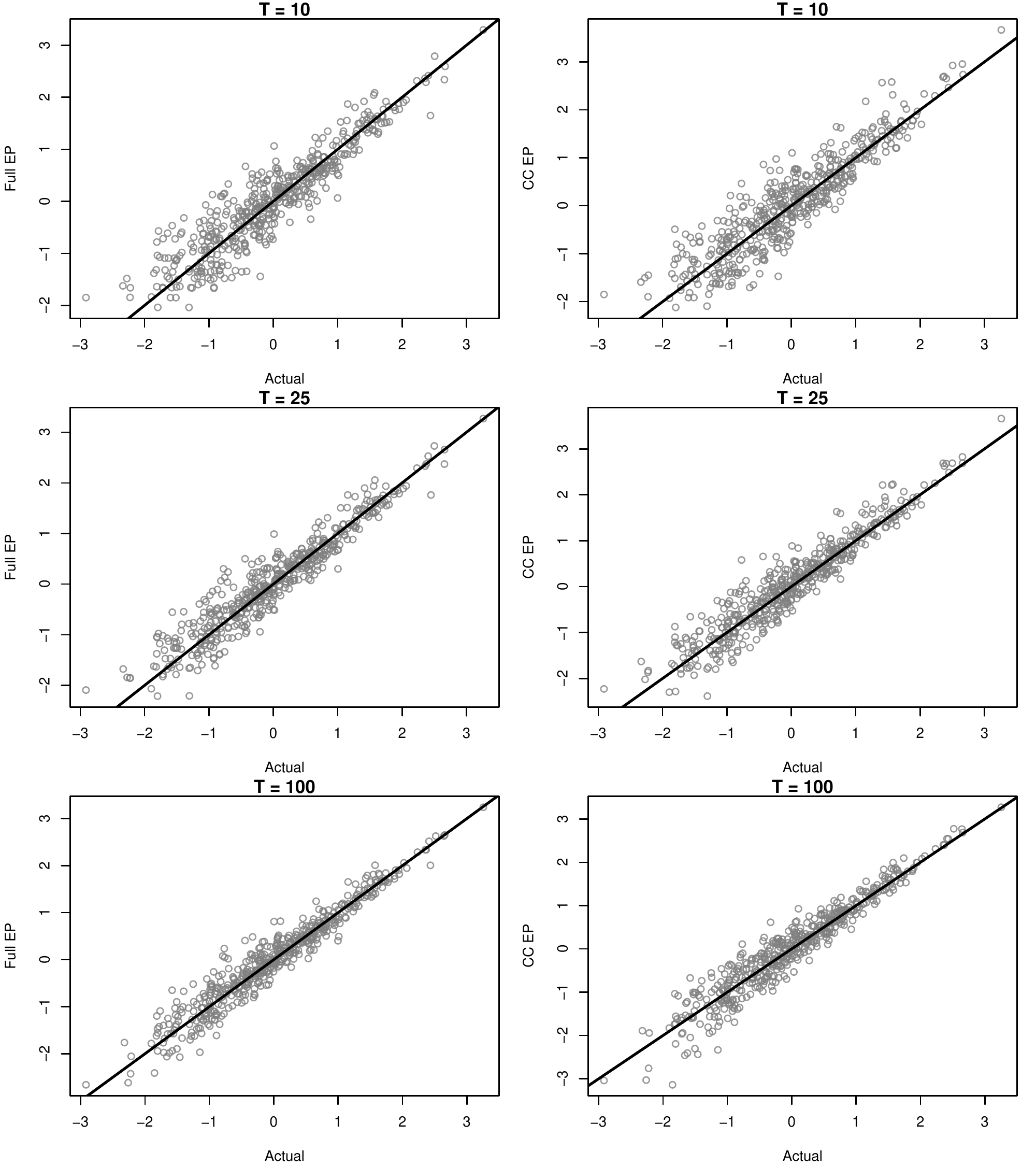}
    \caption{Actual versus estiamted $\alpha_i$ parameter for each node on the logit scale. The left column compares the actual $\alpha_i$ to estimates from the full Power EP, while the right column represents the $\alpha_i$ estimated via the case-control Power EP. }
    \label{fig1:alpha}
\end{figure}

\begin{figure}[ht]
    \centering
    \includegraphics[width=.8\linewidth]{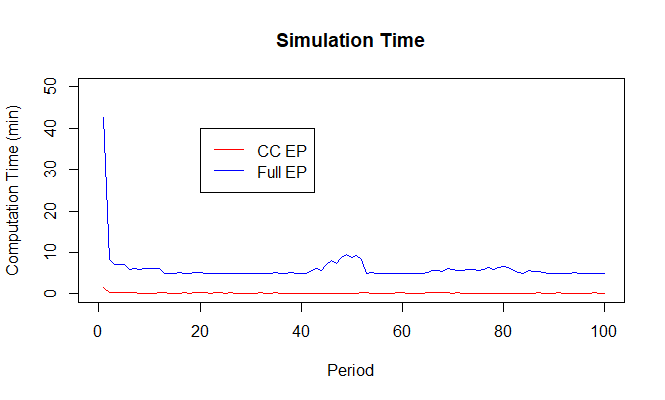}
    \caption{Computation time for the power EP with full data (full EP, blue) and power EP with case-control (CC EP red) method.}
    \label{fig1:simulation_time}
\end{figure}

\end{appendix}
\end{document}